\documentclass[11pt, aps]{revtex4}
\usepackage{graphicx}
\usepackage{amssymb}
\usepackage{amsmath}
\usepackage{latexsym}

\begin{document}

\title{Quantum efficiency of single-photon sources in the cavity-QED \\strong-coupling regime}
\author{Guoqiang Cui and M. G. Raymer}
\affiliation{Oregon Center for Optics and Department of Physics\\
University of Oregon, Eugene, OR 97403 USA}
\email{gcui@uoregon.edu,
raymer@uoregon.edu}

\begin{abstract}
We calculate the integrated-pulse quantum efficiency of
single-photon sources in the cavity quantum electrodynamics (QED)
strong-coupling regime. An analytical expression for the quantum
efficiency is obtained in the Weisskopf-Wigner approximation.
Optimal conditions for a high quantum efficiency and a temporally
localized photon emission rate are examined. We show the condition
under which the earlier result of Law and Kimble [J. Mod. Opt.
\textbf{44}, 2067 (1997)] can be used as the first approximation to
our result.
\end{abstract}

\maketitle


\section{Introduction}

Various implementations of single-photon sources (SPS) based on
atom-like emitters have been reported based on different systems in
the last three decades, such as calcium atoms \cite{Clauser74},
single ions in traps \cite{Diedrich87}, single molecules
\cite{Basche92}, a color center in diamond \cite{Kurtsiefer00}, and
semiconductor nanocrystals \cite{Michler00} or quantum dots (QD)
\cite{Santori01,Yuan02}. The need for efficient single-photon
sources, however, is still a major challenge in the context of
quantum information processing \cite{Bennett92, Knill01}. In order
to efficiently produce single photons on demand, the single quantum
emitter is coupled to a resonant high-finesse optical cavity. A
cavity can channel the spontaneously emitted photons into a
well-defined spatial mode and in a desired direction to improve the
collection efficiency, and can alter the spectral width of the
emission. It can also provide an environment where dissipative
mechanisms are overcome so that a pure-quantum-state emission takes
place. A major question is what is the quantum efficiency (QE) of
the emission from such systems.

Depending on the ratios of the coherent interaction rate $g_0$
between the quantum-emitter and cavity, to the intracavity field
decay rate $2\kappa$, and to the emitter population decay rate
$2\gamma$, we can distinguish two regimes of coupling between the
emitter and the cavity: strong coupling for $g_0>\kappa,\,\gamma$
and weak coupling for $g_0<\kappa,\,\gamma$. The realizations of
cavity-QED strong coupling in the atom-cavity \cite{Kimble94} and
QD-cavity systems \cite{Relthmaier04,Yoshie04,Peter05} allow
researchers to deterministically generate single photons
\cite{McKeever04, Kuhn02}. Single-atom lasers in the strong-coupling
regime have also been studied \cite{Kilin02}. While not in the
strong-coupling regime, Santori et al. \cite{Santori02} showed the
ability to produce largely indistinguishable photons by a
semiconductor QD in a microcavity using a large Purcell effect
\cite{Purcell46}. The QE $\eta_q$ of SPS, which is intrinsic to the
composite quantum system, can be different in these two regimes
because the dynamics of the composite system is different. The
overall efficiency of SPS will also depend on the excitation
efficiency, collection efficiency and detection efficiency, which
are not intrinsic to the composite quantum system; however, they can
be greatly affected by the energy structure of the quantum emitter
and the geometry of the cavity. Qualitative discussions of different
efficiencies based on a particular system in the Purcell
(bad-cavity) regime have been reported in the literature elsewhere
\cite{Santori01}.

In this paper we calculate the integrated-pulse QE of SPS in the
cavity QED strong-coupling regime based on the solutions of the
probability amplitudes in the Weisskopf-Wigner approximation (WWA)
\cite{Weisskopf30}. We find that the QE equals
\begin{equation}
\centering
\eta_q=\left[g_0^{2}/\left(g_0^2+\kappa\gamma\right)\right]
\cdot\left[\kappa/\left(\kappa+\gamma\right)\right].
\label{Efficiency}
\end{equation}
We show the condition under which earlier result associated with Law
and Kimble et al. in \cite{Law97} can be used as the first
approximation to this more complete result. We also establish the
connection between our analytical results and the qualitative
discussions of Pelton et al. in \cite{Pelton02}.

\section{Probability-amplitude method in the Weisskopf-Wigner approximation}

\begin{figure}[htbp]
\centering\includegraphics[width=0.5\textwidth]{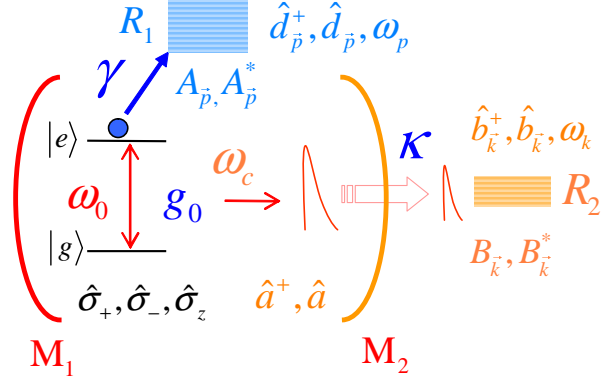}
\caption{Schematic description of a lossy two-level emitter
interacting with a single mode in a leaky optical cavity. $g_0$ is
the coupling constant between the emitter and the cavity field.
$A_{\vec{p}},\,A_{\vec{p}}^{\ast}$ and $B_{\vec{k}},\,
B_{\vec{k}}^{\ast}$ are the coupling constants between the emitter a
single photon and their respective reservoir fields
$\left(R_1,\,R_2\right)$.} \label{f1}
\end{figure}

Consider the interaction of a quantized radiation field with a
two-level emitter located at an antinode of the field in an optical
microcavity, as in Fig. \ref{f1}. $\mathrm{M_1}$ is a perfect
100\%-reflecting mirror and $\mathrm{M_2}$ is a partially
transparent one, from which a sequence of single photons-on-demand
emerges.

The interaction Hamiltonian $\hat{H}_I(t)$ in the interaction
picture for this system in the dipole approximation and
rotating-wave approximation is \cite{Scully97},
\begin{equation}
\centering \hat{H}_I(t)=\hbar
g_0\left(\hat{\sigma}_{+}\hat{a}e^{i\Delta
t}+h.c.\right)+\hbar\sum_{\vec{p}}\left(A_{\vec{p}}^{\ast}\hat{\sigma}_{-}
\hat{d}_{\vec{p}}^{\dag}e^{i\delta_p
t}+h.c.\right)+\hbar\sum_{\vec{k}} \left(B_{\vec{k}}^{\ast}\hat{a}
\hat{b}_{\vec{k}}^{\dag}e^{i\delta_k t}+h.c.\right)
\label{Hamitonian}
\end{equation}
where
$\Delta=\omega_0-\omega_c,\,\delta_p=\omega_p-\omega_0,\,\delta_k=\omega_k-\omega_c
$ are the detunings of the emitter-cavity, emitter-reservoir, and
cavity-reservoir. $\hat{a}$ and $\hat{a}^{\dag}$ are the
annihilation and creation operators for the single cavity mode under
consideration, while $\hat{\sigma}_z$ and $\hat{\sigma}_{\pm}$ are
the Pauli operators for the emitter population inversion, raising,
and lowering, respectively.

At arbitrary time \textit{t}, the state vector can be written as
\begin{eqnarray}
\begin{split}
|\psi(t)\rangle =& E(t)|e,0\rangle|0\rangle_{R_1}|0\rangle_{R_2}+
C(t)|g,1\rangle|0\rangle_{R_1}|0\rangle_{R_2}+ \\
&\sum_{\vec{p}}S_{\vec{p}}(t)|g,0\rangle|1_{\vec{p}}\rangle_{R_1}|0\rangle_{R_2}+
\sum_{\vec{k}}O_{\vec{k}}(t)|g,0\rangle|0\rangle_{R_1}|1_{\vec{k}}\rangle_{R_2}
\end{split}
\label{Wavefunction}
\end{eqnarray}
where $|m,n\rangle \,(m=e,\,g,\;n=0,\,1)$ denotes the emitter state
(excited state, ground state) with $n$ photons in the cavity. $|
j_{\vec{p}}\rangle_{R_1}|l_{\vec{k}}\rangle_{R_2} \,(j,\,l=0,\,1)$
corresponds to $j$ photons in the $\vec{p}$ mode (other than the
privileged cavity mode) of the emitter reservoir $R_1$ and $l$
photons in a single-mode $(\vec{k})$ traveling wave of the
one-dimensional photon reservoir $R_2$ (output beam).
$E(t),\,C(t),\,S_{\vec{p}}(t)$ and $O_{\vec{k}}(t)$ are complex
probability amplitudes.

The equations of motion for the probability amplitudes are obtained
by substituting $|\psi(t)\rangle$ and $\hat{H}_I(t)$ into the
Schr\"odinger equation and then projecting the resulting equations
onto different states respectively. In the WWA \cite{Weisskopf30,
Scully97}, we obtain
\begin{eqnarray}
\dot{E}(t)=-ig_0\exp{(i\Delta t)}C(t)-\gamma E(t),\quad
\dot{C}(t)=-ig_0\exp{(-i\Delta t)}E(t)-\kappa C(t) \label{EOM1} \\
S_{\vec{p}}(t)=-iA_{\vec{p}}^{\ast}\int_0^t dt''\exp{\left(i\delta_p
t''\right)}E(t''),\quad O_{\vec{k}}(t)=-iB_{\vec{k}}^{\ast}\int_0^t
dt'\exp{\left(i\delta_k t'\right)}C(t') \label{EOM2}
\end{eqnarray}
where $\gamma$ and $\kappa$ are one-half the radiative decay rates
of the emitter population (other than the privileged cavity mode)
and the intracavity field, respectively.

Consider the case that the emitter and cavity are at resonance,
$\Delta=\omega_0-\omega_c=0$. By using the initial conditions that
at time $t_0=0$ the quantum emitter is prepared in its excited state
$E(0)=1,\,C(0)=0$ , we obtain the solutions to Eq. (\ref{EOM1}),
\begin{eqnarray}
E(t)&=& \exp(\mathrm{K} t/2)\cdot\left[\cos(gt)+\frac{\Gamma}{2g}\sin(gt)\right] \label{Et} \\
C(t)&=& \exp(\mathrm{K}t/2)\cdot\left[-i\frac{g_0}{g}\sin(gt)\right]
\label{Ct}
\end{eqnarray}
where $\mathrm{K}=\kappa+\gamma,\,\Gamma=\kappa-\gamma$ and
$g\equiv[g_0^2-(\Gamma/2)^2]^{1/2}$ is the generalized vacuum Rabi
frequency. $S_{\vec{p}(t)}$ and $O_{\vec{k}}(t)$ can be obtained by
carrying out the integrations in Eq. (\ref{EOM2}).

\section{Quantum efficiency of SPS in the cavity QED strong-coupling regime}

A single photon will certainly be emitted from the excited emitter,
but it might not be coupled into a single-mode traveling wavepacket
because it can also spontaneously decay to the emitter reservoir.
Define the emission probability $P_o(t)$ to be the probability of
finding a single photon in the output mode of the cavity between the
initial time $t_0=0$ and a later time $t$. This equals
\begin{eqnarray}
P_o(t)=2\kappa\int_0^t
dt'|C(t')|^2=\eta_q\left\{1-\exp(-\mathrm{K}t)\left[1+\frac{\mathrm{K}^2}{2g^2}
\sin^2(gt)+\frac{\mathrm{K}}{2g}\sin(2gt)\right]\right\} \label{Po}
\end{eqnarray}
where $\eta_q$ is given in Eq. (\ref{Efficiency}), by the
single-photon emission probability $P_o(t)$ in the sufficiently
long-time limit $t\gg \mathrm{K}^{-1}$. It may be decomposed as
$\eta_q=\eta_c\cdot\eta_{extr}$, with
\begin{eqnarray}
\eta_c=\frac{g_0^2}{g_0^2+\kappa\gamma}\equiv\frac{2C_0}{2C_0+1},\quad
\eta_{extr}=\frac{\kappa}{\kappa+\gamma} \label{2etas}
\end{eqnarray}
where $C_0\equiv g_0^2/2\kappa\gamma$ is the cooperativity parameter
per emitter \cite{Lugiato84}.

We define $\eta_q$ as the quantum efficiency of SPS in the
cavity-QED strong-coupling regime, which can be viewed as the
product of the coupling efficiency $(\eta_c)$ of the emitter to the
cavity mode and the extraction efficiency $(\eta_{extr})$ of the
single photon into a single-mode traveling wavepacket. The coupling
efficiency characterizes how strong the emitter is coupled to the
privileged cavity mode. The extraction efficiency determines how
large the fraction of light is coupled to a single wave-packet,
outward-traveling-wave mode. We emphasize that the cavity decay is
not considered as a loss, but rather as a coherent out-coupling,
because our goal is to extract single photons from the cavity.

The photon emission rate $n(t)$, defined as the time derivative of
the emission probability, gives the rate of a single photon emerging
from the cavity mirror $\mathrm{M_2}$ and is
\begin{eqnarray}
n(t)\equiv\frac{dP_0(t)}{dt}=2\kappa\frac{g_0^2}{g^2}\exp(-\mathrm{K}
t)\sin^2(gt) \label{rate}
\end{eqnarray}
We expect the shape of $n(t)$ to be sufficiently narrow as to define
a well-localized photon wavepacket and a well-specified time
interval between successively emitted single photons.

\begin{figure}[htbp]
\centering
\begin{minipage}[b]{0.45\textwidth}
\centering
\includegraphics[width=0.95\textwidth]{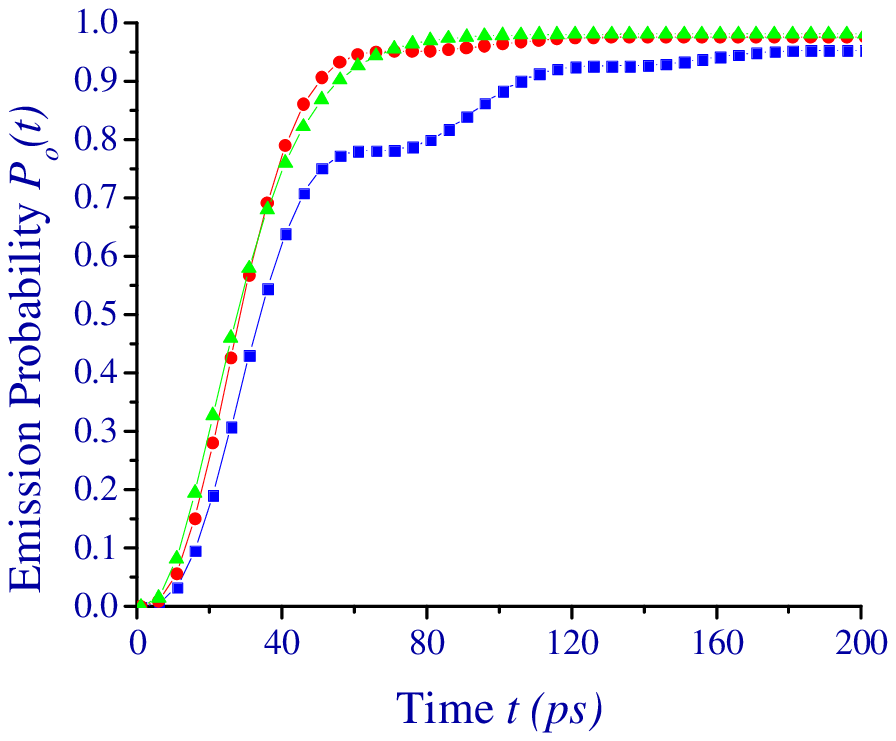}\\
{(a)}
\end{minipage}
\begin{minipage}[b]{0.45\textwidth}
\centering
\includegraphics[width=0.95\textwidth]{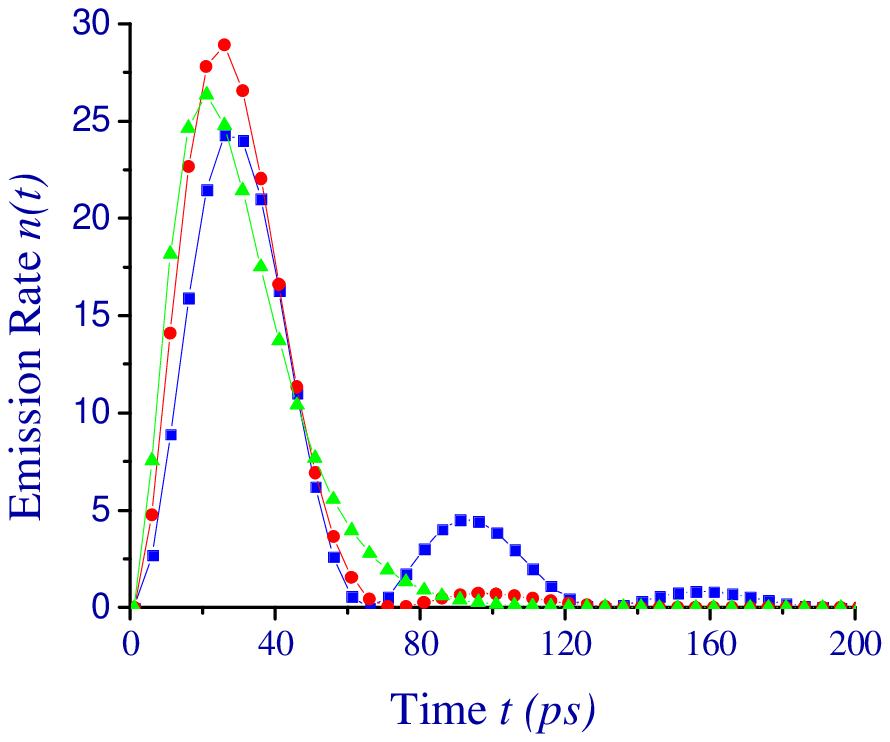}\\
{(b)}
\end{minipage}
\caption{Plots for the time dependence of (a) the emission
probabilities of single photons $P_o(t)$, and (b) the emission rates
$n(t)$, in three different cavity regimes: optimal cavity regime for
$\kappa=g_0^2/\kappa\gg\gamma$, good cavity regime for
$g_0^2/\kappa>\kappa\gg\gamma$, and bad cavity regime for
$\kappa>g_0^2/\kappa\gg\gamma$, (red dot, blue square and green
triangle, respectively) with $\kappa/2\pi=(8.0,\,3.2,\,16)$ GHz,
respectively.} \label{Pn}
\end{figure}

From Eq. (\ref{2etas}), we can see that the larger the ratios
$g_0^2/\kappa\gamma$ and $\kappa/\gamma$, the higher the coupling
efficiency and the extraction efficiency, respectively. For a given
quantum emitter, with no pure dephasing processes, the dipole
dephasing rate is limited by its population decay rate. However, we
can design a cavity with a proper cavity decay rate $\kappa$ to
optimize the QE of SPS and the shape of the photon-emission rate.
Figure \ref{Pn} shows plots of the emission probabilities and the
emission rates for three cavity regimes where we varied the cavity
decay rate $\kappa$, given realistic parameters
$(g_0,\,\gamma)/2\pi=(8.0,\,0.16)$ GHz in each case.

We find that the optimal condition for a high QE and a temporally
narrow emission rate, by optimizing the three parameters in Eq.
(\ref{Efficiency}), is $\kappa=g_0^2/\kappa\gg\gamma$, as shown by
the red dotted curves in Fig. \ref{Pn}. The QE is 96\%, predicted by
Eq. (\ref{Efficiency}) in this example. The photon emission rate is
well localized on the time axis. The width of $n(t)$ is about 32 ps.

\section{Discussion and Conclusion}

An earlier result obtained in the bad-cavity limit by Law and Kimble
is given by \cite{Law97},
\begin{equation}
\centering P(t)\approx\frac{2C_1}{2C_1+1} \label{Law}
\end{equation}
where $C_1=g_0^2/\kappa \gamma_1$ is is the single-atom
cooperativity parameter. Note that the $\gamma_1$ in definition
(\ref{Law}) is the full width of the atomic absorption line. The
cooperativity parameter defined in the present context is $C_0\equiv
g_0^2/2\kappa\gamma$ because here $\gamma$ is the half width, so
these definitions are the same. Comparing our analytical result with
that given by Eq. (\ref{Law}), we see that Eq. (\ref{Law}) is valid
in the limit that spontaneous atomic decay is negligible, as treated
in \cite{Law97}, or equivalently the extraction efficiency
$\eta_{extr}$ is unity. This is not necessary for strong coupling
and is also not implied by the strong-coupling condition. However,
for deterministic production of single photons on demand, we not
only require that the coupling of the emitter to the single cavity
mode is far stronger than its coupling to all other modes
$(g_0^2/\kappa\gg\gamma)$, but also that there needs to be almost no
dephasing of the emitter during the emission process
$(\gamma^{-1}\gg\kappa^{-1})$. This keeps the emission process
deterministic and hence guarantees that the consecutively emitted
photons are indistinguishable.

The Purcell factor, widely referred to in the weak-coupling regime,
is given in \cite{Purcell46} by $F_p=(3\lambda^3/4\pi^2)
\cdot(Q/V)$, which can be shown to be equal to
$F_p=g_0^2/\kappa\gamma_0 \equiv 2C_0\cdot f$, where $\gamma_0$ is
one half the free-space spontaneous decay rate and
$f\equiv\gamma/\gamma_0$ is the fraction of the spontaneous emission
to the modes other than the privileged cavity mode. So our result
for QE can also be written as
\begin{equation}
\eta _q=\frac{F_p}{F_p+f}\cdot\frac{\kappa
}{\kappa+\gamma}=\beta\cdot\frac{\kappa }{\kappa+\gamma}
\label{eta-beta}
\end{equation}
where $\beta \equiv F_p/(F_p+f)$ is called the spontaneous-emission
coupling factor, the fraction of the light emitted by an emitter
that is coupled into one particular mode \cite{Vuckovic02,Lounis05}.
In reference \cite{Pelton02}, the authors discussed the coupling
factor and the extraction efficiency in terms of the quality factor
of the mode. The result Eq. (\ref{eta-beta}) quantifies this
discussion.

To conclude, our result for the QE of SPS in the cavity-QED
strong-coupling regime is more general than earlier results in
\cite{Law97, Pelton02}. It can be used to estimate the QE of single
photons deterministically generated in the cavity output in the
cavity-QED strong-coupling regime, instead of using the Mandel-Q
parameter \cite{McKeever04}. One can improve the QE and performance
of the SPS by optimizing the three parameters in the analytical
result Eq. (\ref{Efficiency}). The QE is crucial for a practical use
of SPS, for example, a high efficiency is required for implementing
the linear-optics quantum computation schemes proposed by Knill et
al. in \cite{Knill01}; while a low efficiency will severely limit
the practical application of SPS in quantum key distribution, as
shown in \cite{Brassard00}.

\section*{Acknowledgement}

This work is supported by NSF under Grant ECS-0323141. We thank
Justin M. Hannigan for discussions.


\end{document}